\documentclass[a4paper,UKenglish]{article}
\pdfoutput=1

\usepackage{upgreek}
\usepackage{scalefnt}
\usepackage{times,mathrsfs}
\usepackage{amsmath,amssymb,bbm,mathtools}
\usepackage[font={small}]{caption}
\usepackage{hyperref}

\hypersetup{
	pdftitle={Quantum linear network coding as one-way quantum computation},
	pdfauthor={Niel de Beaudrap},
	colorlinks,%
	citecolor=black,%
	filecolor=black,%
	linkcolor=black,%
	urlcolor=black
}

\newif\iftikz
\tikzfalse
\tikztrue

\usepackage{amsmath,amssymb,bm,bbm}
\usepackage{enumitem}
\usepackage{tikz}
\usepackage{mathtools}
\usepackage{stmaryrd}

\usetikzlibrary{arrows,decorations.pathmorphing,backgrounds,fit,shapes.geometric,shapes.symbols,calc,positioning}

\newcommand\parStyle[1]{\textrm{\mdseries\upshape({#1}\kern0.1ex)}}

\newlength\romanumlabelwd
\let\le\leqslant

\makeatletter
	\renewcommand\eqref[1]{\textup{\tagform@{\ref{#1}}}}										

	\def\Ddots{\mathinner{\mkern1mu\raise\p@
		\vbox{\kern7\p@\hbox{.}}\mkern2mu
		\raise4\p@\hbox{.}\mkern2mu\raise7\p@\hbox{.}\mkern1mu}}

	\newcommand\eg	{\textit{e.g.}}
	\newcommand\ie	{\textit{i.e.}}
	\newcommand\etc	{\textit{etc}\@ifnextchar.{}{.}}
	\newcommand\etal{\textit{et}~\textit{al}\@ifnextchar.{}{.}}
\makeatother

\DeclareRobustCommand\Z{\mathbb Z}
\DeclareRobustCommand\C{\mathbb C}
\DeclareRobustCommand\idop{\mathbbm 1}

\DeclareMathOperator\GF{GF}

\DeclareMathOperator\img{img}

\newcommand\e{\mathrm e}
\newcommand\x{\times}
\newcommand\ox{\otimes}

\renewcommand\setminus{\smallsetminus}

\newcommand\trans{^{\top\!}}
\newcommand\herm{^\dagger}

\renewcommand\vec[1]{\mathbf{#1}}
\newcommand\unit[1][e]{\vec{\hat #1}}

\newcommand\ket[1]		{\left\lvert#1\right\rangle\mspace{-1.5mu}}		
\newcommand\bra[1]		{\mspace{-1.5mu}\left\langle#1\right\rvert}		
\newcommand\card[1]		{\left\lvert #1 \right\rvert}							

\newcommand\abs\card

\newcommand\mathbox[4]{\mathchoice{\mbox{#1}}{\mbox{#2}}{\mbox{#3}}{\mbox{#4}}}
\newcommand\cZ{\mathbox{$\Lambda\!\!\:{Z}$}{$\Lambda\!\!\:{Z}$}{$\scriptstyle \Lambda\!{Z}$}{$\scriptscriptstyle \Lambda\!{Z}$}}
\newcommand\cX{\mathbox{$\Lambda\!\!\:{X}$}{$\Lambda\!\!\:{X}$}{$\scriptstyle \Lambda\!{X}$}{$\scriptscriptstyle \Lambda\!{X}$}}

\newcommand\sur[1]{^{(#1)}}

\newcommand\toname[2][]   {\xrightarrow[\; #1 \;]{\; #2 \;}}
\newcommand\mapstoname    {\DOTSB\mapstochar\toname}

\newcommand\scalecaps[1]{\text{\scalefont{0.85}#1}}
\newcommand\MBQC{\scalecaps{MBQC}}

\begin{document}

\title{%
	\vspace{-21mm}%
	Quantum linear network coding as one-way quantum computation%
}
\author{%
	\begin{minipage}{0.35\textwidth}\centering
			Niel de Beaudrap\thanks{%
				Supported by a Vidi grant from the Netherlands Organisation for Scientific Research (NWO) and by the European Commission project QALGO.}
			\textsf{\normalsize beaudrap@cwi.nl}\\
			CWI, Amsterdam
	\end{minipage}
	\hspace{15mm}
	\begin{minipage}{0.35\textwidth}\centering
			Martin Roetteler\\
			\textsf{\normalsize martinro@microsoft.com}\\
			Microsoft Research 
	\end{minipage}
	\smallskip
}
\date{1 July 2014}
\maketitle

\vspace{-3ex}

\begin{abstract}
	Network coding~\cite{ACLY00} is a technique to maximize communication rates within a network, in communication protocols for simultaneous multi-party transmission of information.
	Linear network codes are examples of such protocols in which the local computations performed at the nodes in the network are limited to linear transformations of their input data (represented as elements of a ring, such as the integers modulo $2$).
	The quantum linear network coding protocols of Kobayashi~\etal~\cite{KLGNR09b,KLGNR10} coherently simulate classical linear network codes, using supplemental classical communication.
	We demonstrate that these protocols correspond in a natural way to measurement-based quantum computations with graph states over qudits~\cite{RBB03,BB06,DKP07} having a structure directly related to the network.
\end{abstract}


\section{Introduction}
\label{sec:introduction}

\begin{figure}[t]
	\begin{minipage}{0.50\textwidth}
	\caption{%
		The \emph{butterfly network}, with source nodes $S_1$ and $S_2$ and target nodes $T_1$ and $T_2$.
		The two-pair problem on this network is for $S_1$ to communicate their input to the target $T_2$, and simultaneously for $S_2$ to communicate their input to the target $T_1$, assuming that each edge can carry at most one message (represented \eg~by a single bit, $0$ or $1$).
		The classic solution is for $S_1$, $S_2$, and $V_2$ to duplicate their inputs, and for $V_1$, $T_1$, and $T_2$ to compute the parity of their inputs, in which case $(t_1, t_2) = (s_2, s_1)$.
	}
	\label{fig:butterflyTwoPair}
	\end{minipage}
	\hfill
	\begin{minipage}{0.45\textwidth}\centering\vspace*{-2ex}
	\begin{tikzpicture}[scale=0.69,-stealth, line width=1pt, label distance=-3pt, site/.style={circle,draw}]
		\node (null1) at (-0.75,4) {$s_1$};
		\node (null2) at (-0.75,0) {$s_2$};
		\node (S1) at (1,4) [site] {$S_1$};
		\node (S2) at (1,0) [site] {$S_2$};
		\node (V1) at (1.875,2) [site] {$V_1$};
		\node (V2) at (3.625,2) [site] {$V_2$};
		\node (T1) at (4.5,4) [site] {$T_1$};
		\node (T2) at (4.5,0) [site] {$T_2$};
		\node (null3-a) at (6.25,4) {$t_{1}$};
		\node (null4-b) at (6.25,0) {$t_{2}$};

		\foreach \s/\t in {null1/S1, null2/S2, T1/null3-a, T2/null4-b} {%
			\draw (\s) -> (\t);
		}

		\foreach \s/\t/\r/\l/\a in {%
				S1/T1/0.475/m_3/90, S2/T2/0.475/m_5/90, S1/V1/0.5/m_1/180, S2/V1/0.5/m_2/180, V1/V2/0.45/m_4/90, V2/T1/0.4/m_6/0, V2/T2/0.4/m_7/0} {%
			\draw	(\s) -> (\t);
		}
	\end{tikzpicture}
	\end{minipage}
\end{figure}
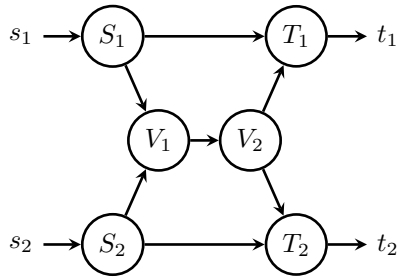

\emph{Network coding}~\cite{ACLY00} is a technique to maximize the rate at which a set of \emph{source nodes} can simultaneously transmit a set of independent messages to certain \emph{target nodes} through a fixed network.
For this purpose, it is sufficient to give each communication link enough bandwidth to accommodate multiple messages to be transmitted at once: however, less bandwidth may be required at each link if one allows nodes to distribute information about the messages across the network.
A classic example is the \emph{two-pair problem} on the ``butterfly network'' (illustrated in Figure~\ref{fig:butterflyTwoPair}): rather than halve the bandwidth between two messages at an apparent bottleneck in the network, the internal nodes may perform simple local computations on the messages, to allow the input data to be reconstructed at the targets.
\emph{Linear network coding} is the special case in which the protocol only requires each node to compute a linear transformation of its inputs to achieve this goal.

We consider \emph{quantum network coding}, in which we perform similar tasks with quantum states transmitted through noiseless quantum channels.
It is immediately apparent that some problems which can be sensibly posed for ``classical'' network coding are impossible in general for quantum network coding.
For instance, while a classical network code allows for the each of the source nodes to each send a copy of their inputs to \emph{both} targets in the butterfly network (see page~\pageref{fig:distribMtxMulticast}), this is clearly not possible for quantum states due to the no-cloning theorem~\cite{NoCloning}.
Other problems which do not require multiple copies of the input states to be re-created at the output (such as the two-pairs problem above) are still potentially unsolvable with fixed-capacity quantum channels alone, even when the corresponding classical problem is solvable~\cite{HINRY06,LOW10}.
However, some of these problems become feasible for quantum states when the network nodes share prior entanglement~\cite{Haya07}, or if the capacities of the communication links scale as the logarithm of the number of target nodes~\cite{SS06}.

Because classical information is easier to faithfully transmit and transform than quantum information, it is common to consider quantum protocols which also allow classical communication, and where fewer restrictions are imposed on the classical than the quantum communication (see Ref.~\cite{NC}).
In a setting where \emph{no} restrictions are imposed on classical communication, Kobayashi \etal~\cite{KLGNR09b} describe a quantum protocol for the \emph{$k$-pairs problem}: the problem in which each of $k$ source nodes wish to communicate their input message to one of $k$ distinct target nodes.
Their protocol is in effect a coherent simulation of a classical linear network code.
More generally, for any classical linear network code which performs some injective linear transformation $\vec t = M \vec s$ of the input data, Ref.~\cite{KLGNR09b} yields a corresponding quantum procedure to coherently simulate that network over for arbitrary superpositions of input data.
We call such a protocol a (classically assisted) \emph{quantum linear network code}.
For the $k$-pairs problem, the protocols of Ref.~\cite{KLGNR09b} were subsequently extended in two different ways by Ref.~\cite{KLGNR10}: to restrict the classical communication to the same network as the quantum communication (albeit with multiple rounds of communication, and sending a single message backwards as well as forwards along each communication link) and to accommodate non-linear protocols as well.

In this article we show that classically assisted quantum linear network codes in the style of Ref.~\cite{KLGNR10} are in effect an instance of \emph{one-way measurement based quantum computation} (\MBQC)~\cite{RBB03,BB06,DKP07,Beau10}: a model of quantum computation in which one may entangle an arbitrary input state $\ket{\psi}$ with a graph state, which is then subjected to a sequence of measurements, leaving a final residual state which contains a transformed state $U\ket{\psi}$ for some unitary transformation\footnote{%
	In general, the transformation which is performed on an input state $\ket\psi$ is not necessarily a unitary transformation, but rather some completely positive trace preserving map $\Phi$ acting on $\rho_0 = \ket{\psi}\bra{\psi}$.
	However, standard treatments of the one-way model describe how measurements on graph states may be used to simulate the transformations performed by unitary circuits, which by construction would transform the input state $\ket{\psi}$ unitarily.}
$U$.
Furthermore, the graph state used as a resource is closely related structurally to the network used in the coding protocol.
This demonstrates a link between \MBQC\ and linear network coding, construed as distributed models of computation, and suggests novel ways of interpreting measurement-based procedures.
At the same time, this suggests \MBQC\ as a unifying framework in which to consider multi-party quantum networking protocols, including cryptographic applications formulated in the one-way model~\cite{BFK09,KMMP09} as well as standard security proofs of BB84~\cite{SP-2000}.




\section{Preliminaries}

In this section, we present introductory remarks on classical linear network coding, and summarize the development of Refs.~\cite{KLGNR09b,KLGNR10}.
We assume familiarity with standard models of quantum computation on qubits, as well as measurement-based quantum computation (see \eg~Refs.~\cite{RBB03,BB06,DKP07,Beau10} for introductory references).
We introduce the notation and the definitions for the operators used over qudits of dimension $d$ below.

\subsection{Classical network coding}
\label{sec:generalNetworkCoding}

\begin{figure}
	\begin{minipage}{0.38\textwidth}
	\caption{%
		An illustration of the transformation of messages performed by a single network node in a linear coding protocol.
	}
	\label{fig:linearNode}
	\end{minipage}
	\hfill
	\begin{minipage}{0.55\textwidth}
	\begin{center}
	\hspace{-1em}
	\begin{tikzpicture}[-stealth, line width=1pt, label distance=-3pt, site/.style={circle,draw}, scale=0.8]
		\node (B1) at (3,0.75) {$b_1$};
		\node (B2) at (3,0.25) {$b_2$};
		\node (B3) at (3,-0.25) {$b_3$};
		\node (B4) at (3,-0.75) {$b_4$};

		\node (M) at (1.5,0) [site] {$M$};

		\node (A1) at (0,0.5) {$a_1$};
		\node (A2) at (0,0) {$a_2$};
		\node (A3) at (0,-0.5) {$a_3$};

		\draw (A1) to (M);
		\draw (A2) to (M);
		\draw (A3) to (M);
		\draw (M) to (B1);
		\draw (M) to (B2);
		\draw (M) to (B3);
		\draw (M) to (B4);


		\node (mtxEqn) [rectangle, fill=gray!15!white, inner sep=2mm] at (6.5,0) {\small$
			\left[\begin{matrix}b_1\\b_2\\b_3\\b_4\end{matrix}\right]
			\!\!\:=\!\!\;
			M\! \left[\begin{matrix}a_1\\a_2\\a_3\end{matrix}\right]
		$};

		\coordinate (Bmid) at ($(B2)!0.5!(B3)$);
		\node () at ($(Bmid)!0.5!(mtxEqn.west)$) {\Large \;$\implies$};

	\end{tikzpicture}
	\end{center}
	\end{minipage}
\end{figure}
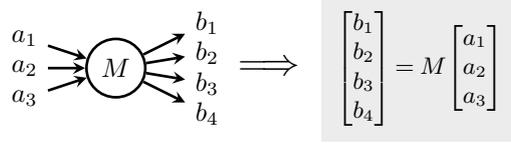

We model a communications network by a directed graph of communications links, each of which can be used to transmit a single message from some message set $M$.
In this article we suppose that $M$ consists of a cyclic ring\footnote{%
	In the setting where messages represent elements of a finite field $\GF(p^r)$ (see \eg~Ref.~\cite{GRB03}), we may replace each communication link with $r$ parallel communications links, representing elements of $\GF(p^r)$ as $r$-dimensional vectors over $\GF(p) \cong \Z_p$.
	In the case of linear network codes, this leads to no loss of generality, as every $\GF(p^r)$-linear transformation of messages is also a $\GF(p)$-linear transformation.}
$\Z_d = \Z/d\Z$.
The messages are sent between co-operative agents (represented by nodes of the digraph) who may perform some non-trivial transformation of the data they receive from ingoing links.
In the context of linear network codes, the transformations performed by each node are linear transformations, as represented in Figure~\ref{fig:linearNode}.
The result of this computation is then sent as output messages to other nodes.
We restrict ourselves to directed acyclic networks, and assume that each node waits for all inputs to arrive before computing its outputs.

The canonical network coding problems involve distributing information from a collection of \emph{source} nodes $\mathsf S = \{S_1, S_2, \ldots \}$ to a collection of \emph{target} nodes $\mathsf T = \{T_1, T_2, \ldots\}$, such as
	the \emph{multicast problem} (in which each source $S_h$ must transmit their data to every one of the targets $T_j$), and 
	the \emph{$k$-pairs problem} (in which each source $S_h$ tries to send their message to a single target $T_{\pi(h)}$, for some permutation $\pi \in \mathfrak S_k$ of the indices).
The source nodes $S_j$ each have some piece of information, usually represented as a single element $s_j \in \Z_d$ or vector $\vec s_j \in \Z_d^{n_j}$.
To put the source and target nodes on an equal footing to the other network nodes, we suppose that the inputs $s_j$ of the sources $S_j$ are messages received from elsewhere (\eg~storage devices owned by the source nodes), and the outputs $t_j$ to be computed by the targets $T_j$ are also transmitted to somewhere, as depicted in Figure~\ref{fig:butterflyTwoPair}.
A solution via linear network codes simply assigns linear transformations to each node, in such a way that the composite transformation performs the correct redistribution of input messages.

\begin{figure}
	\begin{center}
	\hspace*{-2mm}
	\begin{tikzpicture}[scale=0.6,-stealth, line width=1pt, label distance=-3pt, site/.style={circle,draw,font=\small,inner sep=2pt}]
		\node (null1) at (-0.75,4) {$s_1$};
		\node (null2) at (-0.75,0) {$s_2$};
		\node (S1) at (1,4) [site] {$S_1$};
		\node (S2) at (1,0) [site] {$S_2$};
		\node (V1) at (1.875,2) [site] {$V_1$};
		\node (V2) at (3.625,2) [site] {$V_2$};
		\node (T1) at (4.5,4) [site] {$T_1$};
		\node (T2) at (4.5,0) [site] {$T_2$};
		\node (null3-a) at (6.25,4) {$t_{1,1}$};
		\node (null3-b) at (6.25,3.5) {$t_{1,2}$};
		\node (null4-a) at (6.25,0.25) {$t_{2,1}$};
		\node (null4-b) at (6.25,-0.25) {$t_{2,2}$};

		\foreach \s/\t in {null1/S1, null2/S2, T1/null3-a, T1/null3-b, T2/null4-a, T2/null4-b} {%
			\draw (\s) -> (\t);
		}

		\foreach \s/\t/\r/\l/\a in {%
				S1/T1/0.475/m_3/90, S2/T2/0.475/m_5/90, S1/V1/0.5/m_1/180, S2/V1/0.5/m_2/180, V1/V2/0.45/m_4/90, V2/T1/0.4/m_6/0, V2/T2/0.4/m_7/0} {%
			\draw	(\s) -> (\t);
		}

		\node (label) at ($(S2)!0.5!(T2) + (0,-14ex)$)
			{\small
				\begin{minipage}{15mm}\footnotesize\itshape
					Multicast \\ problem on \\ the butterfly \\ network:
				\end{minipage}\!
			$
				\left[\begin{matrix}t_{1,1} \\ t_{1,2} \\ t_{2,1} \\ t_{2,2} \end{matrix}\right]
				\!\!\;=
				\left[\begin{matrix} 1 & 0 \\ 0 & 1 \\ 1 & 0 \\ 0 & 1 \end{matrix}\right]
				\left[\begin{matrix}s_1\\s_2\end{matrix}\right]
			$
	};
	
		\node (problem) [rectangle, fill=gray!15!white, inner xsep=2mm] at (13.2,0.5) {%
			\begin{minipage}{60mm}%
			\vspace*{-2mm}\small
			\setlength{\arraycolsep}{2.5pt}%
			\begin{align*}
		\text{Decompose:}\mspace{-36mu}\\
				\left[\begin{matrix} 1 & 0 \\ 0 & 1 \\ 1 & 0 \\ 0 & 1 \end{matrix}\right]
			&=
			\left[\begin{matrix} T_1 & 0 \\ 0 & T_2 \end{matrix}\right]
			\left[\begin{matrix} 1 & 0 & 0 \\ 0 & V_2 & 0 \\ 0 & 0 & 1 \end{matrix}\right]
			\left[\begin{matrix} 1 & 0 & 0 \\ 0 & V_1 & 0 \\ 0 & 0 & 1 \end{matrix}\right]
			\left[\begin{matrix} S_1 & 0 \\ 0 & S_2 \end{matrix}\right] \\
			& \phantom{=} \text{where } \begin{aligned}[t]
													S_1, S_2 &: \Z_d \to \Z_d^2		\\[-0.75ex]
													V_1 &: \Z_d^2 \to \Z_d 				\\[-0.75ex]
													V_2 &: \Z_d \to \Z_d^2 				\\[-0.75ex]
													T_1, T_2 &: \Z_d^2 \to \Z_d^2
			                \end{aligned}
			\end{align*}
			\end{minipage}};

	\end{tikzpicture}
	\end{center}
	\vspace{-1ex}
	\caption{%
		The multicast problem on the butterfly network, formulated as a linear transformation over the ring $\Z_d$.
		A solution by linear network coding decomposes this transformation as a product of block matrices according to the network structure.
		A typical solution to this problem is presented in Eqn.~\eqref{eqn:butterflyMulticastSoln}.
	}
\label{fig:distribMtxMulticast}
\end{figure}
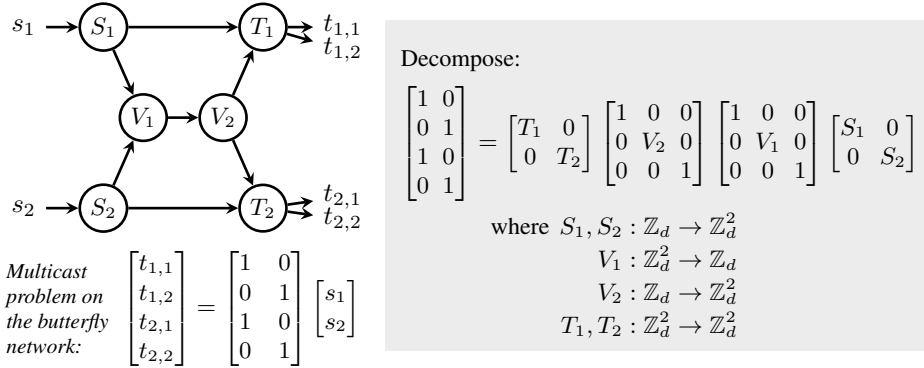
We regard linear network coding as a distributed model of computation, in which linear transformations are decomposed into block matrices, where each non-trivial block is represented by a single node.
For \emph{any} linear function $f$ --- of which the $k$-pairs and multicast problems are special cases --- we consider which transformations the nodes may perform (if any) to compute $f$.
Figure~\ref{fig:distribMtxMulticast} presents the multicast problem on the butterfly network in this form, to which one solution is the following assignment of matrices to each node in the network:
  \begin{align}
	\label{eqn:butterflyMulticastSoln}
 		S_1 = S_2 &= V_2 = \left[\begin{matrix}1 \\ 1\end{matrix}\right]\!,\!
	&
		V_1 &= \left[\begin{matrix}\,1\; & \;1\,\end{matrix}\right]\!,\!
&
		T_1 \!\!\!\:&\,=
			\left[\!\!\:\begin{matrix} \phantom{-}1 \:&\: 0 \\ -1 \:&\: 1	\end{matrix}\;\right]\!,\!
	&
		T_2 &=
			\left[\;\begin{matrix} 1 \:& -1 \\ 0 \:& \phantom{-}1	\end{matrix}\;\right]\!.    
  \end{align}

\subsection{Classically assisted quantum network coding}
\label{sec:coherentNetworkCoding}

We now outline the constructions of Ref.~\cite{KLGNR09b}, and also of Ref.~\cite{KLGNR10} in the special case of linear coding protocols over the ring $\Z_d$ of integers modulo $d$, for protocols using message qudits of dimension $d$.

%
Consider a node $V$ performing some coding operation $\vec y = V \vec x$ for $\vec x \in \Z_d^\ell$ and $\vec y \in \Z_d^m$ in a classical coding network.
We may simulate this node by initializing an output register $\vec y = \vec 0 \in \Z_d^m$, performing a bijective mapping $(\vec x, \vec y) \mapsto (\vec x,\, \vec y + V \vec x)$ in the larger space $\Z_d^{\:\!\ell + m}$\!, and then discarding the input $\vec x$.
The bijective mapping can be performed by elementary row transformations on $\vec x$, which in the quantum setting may be performed by controlled-$X$ operations,
\begin{equation}
		\cX_{j,k}	=	\sum_{c = 0}^{d-1}	\ket{c}\bra{c}_j	\ox X^{\:\!c}_{\!\!\:k},
\end{equation}
where $X \ket{q} = \ket{q + 1 \bmod{d}}$ is an analogue of the unitary Pauli operator $\sigma_x$ on qubits.
Consider a generic node $V$ which accepts a collection of input qudits $a_1, \ldots, a_\ell$ as input and produces output qudits $b_1, \ldots, b_m$, coherently simulating the transformation
$
		\ket{\vec x}_{\!a_1\cdots a_\ell} \longmapsto \ket{T\vec x}_{\!b_1 \cdots b_m}
$. 
In the construction of Ref.~\cite{KLGNR09b} for quantum linear codes, $V$ simulates this transformation by preparing the qudits $b_1, \ldots, b_k$ in the $\ket{0}$ state, and performing the transformations
\begin{equation}
		\cX^{V_{\!\!\;j,k}}	\Bigl( \ket{x_k} \ox \ket{0} \Bigr)
	=
		\ket{x_k} \ox \ket{V_{j,k} x_k}
\end{equation}
on the qudits $a_k$ and $b_j$, for every index $1 \le j \le \ell$ and $1 \le k \le m$ in any order.
For standard basis states, the result is to transform $\ket{\vec x} \ket{\vec 0} \mapsto \ket{\vec x} \ket{V \vec x}$.
This characterizes a linear transformation
\begin{equation}
	\label{eqn:coherentLinearEmbedding}
		\tilde U_V
	\;=\;
		\left( \prod_{j=1}^m \prod_{k=1}^\ell \cX_{a_k,b_j}^{V_{\!\!\;j,k}}	\right) \Biggl( \idop_{\vec a} \ox \ket{\vec 0}_{\vec b}\Biggr)\;,
\end{equation}
which is a unitary embedding for any transformation $V$.
(An example of such a circuit is illustrated in Figure~\ref{fig:decompose-U_V}
.)
If the qudits $a_1, \ldots, a_\ell$ where originally in standard basis states, we could simply discard them; but if they are initially not in standard basis states, they will become entangled with $b_1, \ldots, b_m$.
To decouple them, we attempt to project each of the qudits $a_j$ to the $\ket{+}$ state by measurement,
\begin{equation}
	\label{eqn:+state}
		\ket{+}	\;=\;	\tfrac{1}{\sqrt d}\Bigl( \ket{0} + \ket{1} + \cdots + \ket{d-1} \Bigr).
\end{equation}
Successfully doing so on a generic input state $\ket{\psi} = \sum_{\vec x} u_{\vec x} \ket{\vec x}$ would lead to the sequence of transformations
\begin{equation}
	\label{eqn:coherentCodingNodeAction}
	\begin{aligned}[b]
		\ket{\psi}
	\;
	\longmapsto
		\sum_{\vec x} u_{\vec x} \ket{\vec x}_{\!\vec a}\ket{\vec 0}_{\!\vec b}
	\;&\longmapsto
		\sum_{\vec x} u_{\vec x} \ket{\vec x}_{\!\vec a}\ket{V \vec x}_{\!\vec b}
	\\&\longmapsto
		\frac{1}{\sqrt{d^{\ell}}} \left(\bigotimes_{k = 1}^\ell \ket{+}_{a_k} \!\right) \ox 
		\sum_{\vec x} u_{\vec x} \ket{V \vec x}_{\!\vec b}	\,.\,
	\end{aligned}
\end{equation}
This mapping is of course non-unitary:
projection onto $\ket{+}$ must be performed as part of a measurement onto some basis.
Ref.~\cite{KLGNR09b} considers a measurement of the qudits $a_j$ in the Fourier basis,
\begin{align}
		\ket{\omega_r}
	\,&=
		\frac{\mbox{\small $1$}}{\mbox{\small $\sqrt d$}} \sum_{x = 0}^{d-1} \e^{2 \pi i x r/d} \ket{x}
	\,=
		F \ket{r},
	&
	\text{where}~
	F &= \frac{\mbox{\small $1$}}{\mbox{\small $\sqrt d$}} \!\sum_{x,r = 0}^{d-1} \!\e^{2 \pi i k x/d} \ket{x}\bra{r}.
\end{align}
The operator $F$ is the \emph{quantum Fourier transform over $\Z_d$}.
We may attempt to simulate projection of each qudit $a_j$ onto $\ket{+}$ by Fourier basis measurements, where a result of $\ket{\omega_0}$ is a success, as $\ket{\omega_0} = \ket{+}$.
If we obtain results $\ket{\omega_{r_j}}$ for $r_j \ne 0$ instead of $\ket{+}$, the post-measurement state is 
\begin{align}
		\left(\bigotimes_{k = 1}^\ell \ket{\omega_r}_{a_k} \!\right) \ox 
		\sum_{\vec x} u_{\vec x} \e^{-2 \pi i (\vec r \cdot \vec x)/d} \ket{V \vec x}_{\!\vec b}
\end{align}
up to normalization.
If $V$ is injective, the relative phase $\e^{-2\pi i (\vec r \cdot \vec x)/d}$ can be undone by a suitable application of $Z$ operations on the qudits $b_1, \ldots, b_m$, where $Z$ is the unitary generalization of $\sigma_z$:
\begin{equation}
		Z	\;=\;	\sum_{q = 0}^{d-1} \e^{2 \pi i q / d} \ket{q}\bra{q}	.
\end{equation}
If $V$ is not injective, then only certain vectors $\vec r$ of measurement outcomes can be immediately corrected, resulting in a non-unitary CP map.
However, regardless of whether some nodes in coding network perform non-invertible operations, the relative phases which accumulate on the entire state are linear functions.
Then if the transformation performed by the whole network is injective, the phases which have accumulated due to the measurements can be undone if the target nodes have sufficient information about the measurement outcomes.

The protocol of Ref.~\cite{KLGNR09b} solves the $k$-pairs problem: thus the transformation it performs is indeed injective.
Each node simply transmits their measurement outcomes to each target node, which performs a suitable combination of $Z$ operations to correct the relative phases.
Ref.~\cite{KLGNR10} presents an alternative protocol in which the measurements are deferred until after all quantum messages have been sent, and in which the internal nodes of the network do the majority of the phase corrections, as follows.
Consider a node which attempts to coherently simulate a transformation $L: \Z_d^\ell \to \Z_d^m$ in the middle of a coding network which attempts to coherently simulate a transformation $M: \Z_d^{\mathscr S} \to \Z_d^{\mathscr T}$ on an input state $\ket{\psi} = \sum_{\vec x} u_{\vec x} \ket{\vec x}$.
Suppose that we perform the simulation procedure above, but omitting the Fourier basis measurements. 
For some linear maps $H$ and $K$, the state after the final quantum messages is in general an entangled state of the form\footnote{%
	The final tensor factor is on the remaining nodes entangled with the sources, whose components in the standard basis are again some linear transformations of the standard basis on the source nodes' inputs; by induction on the depth of the coding network, one may show that $H$ and $K$ are indeed linear transformations.
}
\begin{equation}
\label{eqn:entangledQuantumNetworkCode}
\mspace{-10mu}
\begin{aligned}[b]
		\ket{\Psi}
	\,=\,
		\sum_{\vec x} \!\!\; u_{\vec x} \ket{\vec x}_{\!\mathsf S}
															&\ox \ket{M \vec x}_{\!\mathsf T} 
														\ox \Bigl( \ket{K \vec x}_{\!a_1, \ldots, a_\ell} \ox \ket{LK \vec x}_{\!b_1, \ldots, b_m} \Bigr)
														\ox \ket{H \vec x}_{\!\text{rest}} \,,
\end{aligned}\mspace{-50mu}
\end{equation}
where the factors in parentheses are the input and output qudits to the node $L$.
If the qudits $b_1, \ldots, b_m$ are measured in the Fourier basis by the nodes to which they are sent, they yield some outcomes $r_1, \ldots, r_m$, and the remaining qudits are transformed to 
\begin{equation}
\label{eqn:measOnEntangledQuantumNetworkCode}
\mspace{-10mu}
\begin{aligned}[b]
		\ket{\Psi'}
	\,=\,
		\sum_{\vec x} \!\!\; u_{\vec x} \ket{\vec x}_{\!\mathsf S}
															&\ox \ket{M \vec x}_{\!\mathsf T} 
														\ox \Bigl( \e^{-2\pi i (\vec r \cdot LK \vec x)/d} \ket{K \vec x}_{\!a_1, \ldots, a_\ell} \Bigr)
														\ox \ket{H \vec x}_{\!\text{rest}} \,,
\end{aligned}\mspace{-50mu}
\end{equation}
where $\vec r$ is the vector of the outcomes.
Let $\bm \tau = L\trans \vec r$: we have $\bm \tau \cdot K\vec x = \vec r \cdot LK\vec x$ by construction.
If the nodes which perform these measurements send the outcomes to the node $L$, then $L$ can undo the phases induced by measurement of the qudits $b_k$ by performing the operation $Z^{\bm \tau} := Z_{a_1}^{\tau_1} Z_{a_2}^{\tau_2} \cdots Z_{a_\ell}^{\tau_\ell}$, which performs the mapping
\begin{equation}
\begin{aligned}[b]
		Z_{a_1}^{\tau_1} Z_{a_2}^{\tau_2} \cdots & Z_{a_\ell}^{\tau_\ell} \ket{\Big.\big(K\vec x\big)_{\!1} \big(K\vec x\big)_{\!2} \cdots \big(K \vec x\big)_{\!\ell}}
	\\[0.5ex]&=\;
		\exp\Bigl(\tfrac{2\pi i}{d}\bigr[\tau_1 (K \vec x)_1 + 
			\cdots + \tau_\ell (K \vec x)_\ell \bigr]\Bigr) \ket{K \vec x}
	\\[1ex]&=\;
		\e^{2 \pi i (\bm \tau \cdot K \vec x)/d} \ket{K \vec x}.
\end{aligned}
\end{equation}
Performing these corrections on $\ket{\Psi'}$ then yields the state
\begin{equation}
\mspace{-10mu}
\begin{aligned}[b]
		\ket{\Psi''}
	\,=\,
		\sum_{\vec x} \!\!\; u_{\vec x} \ket{\vec x}_{\!\mathsf S}
															&\ox \ket{M \vec x}_{\!\mathsf T} 
														\ox \ket{K \vec x}_{\!a_1, \ldots, a_\ell} 
														\ox \ket{H \vec x}_{\!\text{rest}} \,,
\end{aligned}
\end{equation}
which has fewer unmeasured qudits than $\ket{\Psi}$, and no relative phases.
This simulates projecting the qudits $b_1, \ldots, b_m$ to the $\ket{+}$ state.
By induction, if each node aside from the source nodes (but including the target nodes) measures their input qudits in the Fourier basis, and communicates the outcomes backwards along their incoming links to the nodes which provided those qudits, those nodes can correct for the effect of the measurements.
Eventually one obtains the state
\begin{equation}
		\ket{\smash{\Psi^{(n)}}\big.}
	\;\!=\;\!
		\sum_{\vec x} \!\!\;u_{\vec x} \ket{\vec x}_{\!\mathsf S} \ox \ket{M \vec x}_{\!\mathsf T},
\end{equation}
which is an entangled state of the (collective) inputs to the source nodes and the outputs of the target nodes.
If the source nodes measure their qudits in the Fourier basis, it suffices for them to communicate the outcomes to target nodes in such a way that the outcomes can be corrected.

For arbitrary linear transformations $M$, direct communication among target nodes or between the source and the target nodes may be required to undo the relative phases induced by measurement.
If the source nodes measure their qudits and collectively obtain a vector $\vec s$ of outcomes, the resulting state on the remaining qudits is
\begin{equation}
		\ket{\smash{\Psi^{(n+1)}}\big.}
	\;\!=\;\!
		\sum_{\vec x} \!\!\;u_{\vec x} \e^{-2\pi i (\vec s \cdot \vec x)/d} \ket{M \vec x}_{\!\mathsf T}.
\end{equation}
If $M$ has a left-inverse $A$, and we let $B = A\trans$, it suffices for the sources to somehow communicate $\sigma_j := \sum_k B_{jk} s_k$ to the target node $T$ which is responsible for producing the message $t_j$.
This would allow $T$ to perform a $Z^{\sigma_{\!j}}$ correction and undo the relative phase on the $j\textsuperscript{th}$ output qudit.
Specifically, if the sources collectively communicate $\bm \sigma = B \vec s$ to the targets, who collectively perform the phase operations $Z^{\bm \sigma} = Z^{\sigma_1}_{t_1} Z^{\sigma_2}_{t_2} \cdots$ on the target qudits, the resulting state is
\begin{equation}
	\!\!
	\begin{aligned}[b]
		\ket{\smash{\Psi^{(n+2)}}\big.}
	\;=\;
		\sum_{\vec x} \!\!\;u_{\vec x} \e^{2\pi i \bigl[\bm \sigma \cdot (M\vec x) - \vec s \cdot \vec x\bigr]/d} \ket{M \vec x}_{\!\mathsf T}
	\;&=\;
			\sum_{\vec x} \!\!\;u_{\vec x} \e^{2\pi i [\vec s\!\trans \!\!\:(A M - \idop) \vec x]/d} \ket{M \vec x}_{\!\mathsf T}
	\\&=\;
			\sum_{\vec x} \!\!\;u_{\vec x} \ket{M \vec x}_{\!\mathsf T};
	\end{aligned}
	\mspace{-39mu}
\end{equation}
There are special cases where the amount of communication required outside of the network can be bounded.
In particular, for the $k$-pairs problem where $M$ is a permutation matrix (so that $(M^{-1})^\top = M$), it suffices to perform the classical linear coding protocol on the vector $\vec s$ to transmit $\bm \sigma = M \vec s$ to the target nodes.
In this case, all classical communications may be restricted to the same network as the quantum communications --- albeit using each communication link once in reverse, for the measurements of the qudits involved in the intermediate messages.
More generally, if $M$ is injective and there is a block-diagonal matrix $B$ (where the blocks act on collections of messages held by individual target nodes) such that $M\trans B M = \idop$, the sources may communicate $M \vec s$ to the targets, allowing the target nodes to compute $\bm \sigma = B\trans M \vec s$ and use this to govern phase corrections.


\section{Classically assisted quantum linear coding \\[-0.25ex] is one-way \MBQC}
\label{sec:qmLinearCodingAsMBQC}

We now show how any coherent linear coding protocol, as described in Section~\ref{sec:coherentNetworkCoding}, is in essence a measurement computation in the one-way model.
The graph states of the \MBQC\ procedures constructed in this way are easily derived from the coding network itself: allocate two entangled qudits at either end of each communications link in the network (one for the node on either side of the link), with further entangling operations between the qudits corresponding to the incoming links and the outgoing links.
The corrections are the same as for the coherent coding network, albeit with some supplemental corrections arising from the way that the $\cX$ operations are simulated.
If we follow the protocol of Ref.~\cite{KLGNR09b}, the corrections are all deferred to the end of the procedure, as in standard treatments of measurement-based computation.

Again, we assume familiarity with the measurement based model: see Refs.~\cite{RBB03,DKP06,BB06,Beau10} for references applicable to qubits (similar results and constructions apply over arbitrary qudits).

\subsection{\MBQC\ simulation of a single coding node}

\begin{figure}
\begin{center}
\begin{minipage}{0.225\textwidth}
\begin{tikzpicture}[scale=0.28, label distance=0pt]
  \node [label=left:$\ket{x_1}\!\!\!\!$] (x1) at (-0.5,2) {};
	\node [label=left:$\ket{x_2}\!\!\!\!$] (x2) at (-0.5,4) {};
	\node [label=left:$\ket{x_\ell}\!\!\!\!$] (xl) at (-0.5,8) {};
	\node [label=left:$\ket{0}\!\!\!\!$] (y0) at (-0.5,0) {};
	
	\coordinate (x1') at (1,2);
	\coordinate (x2') at (2,4);
	\coordinate (xl') at (4,8);

	\coordinate (y1) at (1,0);
	\coordinate (y2) at (2,0);
	\coordinate (yl) at (4,0);
	
	\node (x1'') at (5.5,2) [label=right:$\!\!\!\!\ket{x_1}$] {};
	\node (x2'') at (5.5,4) [label=right:$\!\!\!\!\ket{x_2}$] {};
	\node (xl'') at (5.5,8) [label=right:$\!\!\!\!\ket{x_\ell}$] {};
	\node (yf) at (5.5,0) [label=right:$\!\!\!\!\ket{\vec v \!\cdot\! \vec x}$] {};

	\filldraw [black] (x1') circle (7pt);
	\filldraw [black] (x2') circle (7pt);
	\filldraw [black] (xl') circle (7pt);

	\filldraw [black] ($(x2')!0.4!(xl')$) circle (2pt);
	\filldraw [black] ($(x2')!0.5!(xl')$) circle (2pt);
	\filldraw [black] ($(x2')!0.6!(xl')$) circle (2pt);

	\filldraw [black] (2.6,1) circle (2pt);
	\filldraw [black] (3,1) circle (2pt);
	\filldraw [black] (3.4,1) circle (2pt);

	\draw [black] (y1) circle (13pt) node [label=below:\small$v_{\!\!\;1}$] {};
	\draw [black] (y2) circle (13pt) node [label=below:\small$v_{\!\!\;2}$] {};
	\draw [black] (yl) circle (13pt) node [label=below:\small$v_{\!\!\;\ell}$] {};

	\draw [black] (x1') to ($(y1) + (0,-13pt)$);
	\draw [black] (x2') to ($(y2) + (0,-13pt)$);
	\draw [black] (xl') to ($(yl) + (0,-13pt)$);

	\draw [black] (x1) to (x1'');
	\draw [black] (x2) to (x2'');
	\draw [black] (xl) to (xl'');
	\draw [black]	(y0) to (yf);

\end{tikzpicture}
\end{minipage}
\;\;\;\;\;\;{\large$\equiv$}\,\;
\begin{minipage}{0.225\textwidth}
\begin{tikzpicture}[scale=0.28, label distance=0pt]
  \node [label=left:$\ket{x_1}\!\!\!\!$] (x1) at (-0.25,2) {};
	\node [label=left:$\ket{x_2}\!\!\!\!$] (x2) at (-0.25,4) {};
	\node [label=left:$\ket{x_\ell}\!\!\!\!$] (xl) at (-0.25,8) {};
	\node [label=left:$\ket{+}\!\!\!\!$] (y0) at (-0.25,0) {};
	
	\coordinate (x1') at (1,2);
	\coordinate (x2') at (2,4);
	\coordinate (xl') at (4,8);

	\coordinate (y1) at (1,0);
	\coordinate (y2) at (2,0);
	\coordinate (yl) at (4,0);
	\node [draw, inner sep=2pt] (yF) at (5.5,0) {\footnotesize$F\herm$};
	
	\node (x1'') at (7.4,2) [label=right:$\!\!\!\!\ket{x_1}$] {};
	\node (x2'') at (7.4,4) [label=right:$\!\!\!\!\ket{x_2}$] {};
	\node (xl'') at (7.4,8) [label=right:$\!\!\!\!\ket{x_\ell}$] {};
	\node (yf) at (7.4,0) [label=right:$\!\!\!\!\ket{\vec v \!\cdot\! \vec x}$] {};

	\filldraw [black] (x1') circle (7pt);
	\filldraw [black] (x2') circle (7pt);
	\filldraw [black] (xl') circle (7pt);

	\filldraw [black] ($(x2')!0.4!(xl')$) circle (2pt);
	\filldraw [black] ($(x2')!0.5!(xl')$) circle (2pt);
	\filldraw [black] ($(x2')!0.6!(xl')$) circle (2pt);

	\filldraw [black] (2.6,1) circle (2pt);
	\filldraw [black] (3,1) circle (2pt);
	\filldraw [black] (3.4,1) circle (2pt);

	\filldraw [black] (y1) circle (7pt) node [label=below:\small$v_{\!\!\;1}$] {};
	\filldraw [black] (y2) circle (7pt) node [label=below:\small$v_{\!\!\;2}$] {};
	\filldraw [black] (yl) circle (7pt) node [label=below:\small$v_{\!\!\;\ell}$] {};

	\draw [black] (x1') to (y1);
	\draw [black] (x2') to (y2);
	\draw [black] (xl') to (yl);

	\draw [black] (x1) to (x1'');
	\draw [black] (x2) to (x2'');
	\draw [black] (xl) to (xl'');
	\draw [black]	(y0) to (yF) to (yf);
\end{tikzpicture}
\end{minipage}
\end{center}
\vspace*{-1ex}
\caption{%
	Equivalent ways to decompose a unitary transformation $\tilde U_V$ which prepares a single message qudit, for a single-row matrix $V = \vec v\trans$.
	The left-hand circuit represents the decomposition of Eqn.~\eqref{eqn:coherentLinearEmbedding}. 
	Variables $v_j$ below operations denote the power to which the circuit operation is raised.
	Multi-row coding transformations $V$ may be simulated by several such circuits, acting on different target qudits.
}
\label{fig:decompose-U_V}
\end{figure}
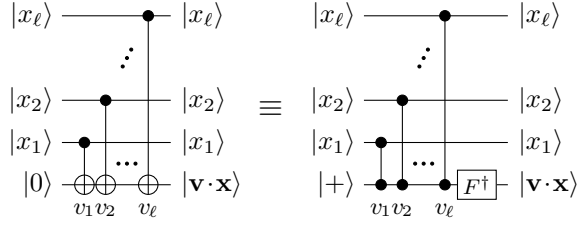

The main element of the correspondence between quantum linear network coding and \MBQC\ is the observation that $\cX$ operations differ by only a Fourier transform from a controlled-phase operation,
\begin{equation}
  \cZ \,=\,	(\idop \ox F) \cX (\idop \ox F\herm) \,=\, \sum_{c = 0}^{d-1}	\ket{c}\bra{c}	\ox Z^{\:\!c}\!,
\end{equation}
which are the diagonal operations used to construct the entanglement structures in measurement-based computation.
This means that the injective maps $\tilde U_V$ used to perform the coding at each node may be straightforwardly represented in terms of preparing the state $\ket{+} = F\ket{0}$ for each output qudit $b_j$ to be sent, performing the entangling operation $\cZ^{V_{j,k}}$ between each input qudit $a_k$ and each output qudit $b_j$, and then acting on $b_j$ with a Fourier transform, as represented in Figure~\ref{fig:decompose-U_V}.

Note that the inverse Fourier transform acting on the output-message qudit may be simulated by a Fourier basis measurement by introducing another auxiliary qudit, using a standard \MBQC\ construction.
Consider a qudit $v$ in an arbitrary pure state $\ket{\psi} = \sum_{x=0}^{d-1} u_x \ket{x}$.
We may introduce a qudit $w$ prepared in the state $\ket{+}$, and entangle them using a $\cZ\herm$ operation, obtaining the state
\begin{equation}
  \ket{\Psi}_{\!vw} = \cZ_{vw}\herm \ket{\psi}_{\!v} \ket{+}_{\!w} \,.
\end{equation}
We then measure $v$ in the Fourier basis, obtaining a state $\ket{\omega_r}$, and perform the operation $X^{-r}$ on $w$.
We may use the stabilizer formalism (see \eg~Ref.~\cite{Beaudrap2013b}) to succinctly verify how this sequence of transformations, considered as CP maps, transform $X$ and $Z$: as these generate an operator basis for single-qudit states, this will suffice to show how it transforms $\ket{\psi}_v$ to $F\herm \ket{\psi}_w$.
Specifically, we wish to see how the group of Pauli operators which \emph{stabilize} the state (\ie,~at each point in time, those Pauli operators for which the state is a $+1$-eigenvector) transforms, for states on $v$ and/or $w$.
We use the following facts:
\begin{itemize}
\item 
	We write $\omega = \exp(\frac{2\pi i}{d}) \in \C$ as a minor abuse of notation: it is easy to verify that $X \ket{\omega_r} = \omega^r \ket{\omega_r}$.
	In particular, $\ket{+}$ is the unique $+1$-eigenvector of $X$ up to scalar factors.
\item
	Measuring $v$ in the Fourier basis is equivalent to measuring the eigenstates of $X_v$, obtaining some state $\ket{\omega_r}$: the post-measurement state is then stabilized by $\omega^{-r} X_v$, as well as by operators (but only those operators) which commute with $X_v$ and stabilized the pre-measurement state.
\item
	Conjugating $X_v$ by $\cZ_{vw}\herm$ yields $X_v Z_w\herm$, and similarly conjugating $X_w$ by $\cZ_{vw}$ yields $Z_v\herm X_w$.
	As they are diagonal, conjugating $Z_v$ or $Z_w$ by $\cZ_{vw}$ has no effect.
	Conjugating by $X_w^{-r}$ transforms $Z_w\herm$ to $\omega^{-r} Z\herm$, and leaves $X_w$ unchanged.
\end{itemize}
We may then describe the sequence of transformations on $\ket{\psi}_v$ as follows: for any scalar $\phi \in \C$, the operator $\phi X_v$ transforms as follows:
\begin{subequations}
\begin{equation}
  \begin{aligned}[b]
			\left\langle \phi X_v \right\rangle
		\;\mapstoname{\text{prep.\ } \ket{+}_w}&
			\left\langle \begin{matrix} \phi X_v \,,\, X_w \end{matrix}\right\rangle
		\\[-1.25ex]\mapstoname{\cZ_{\!vw}\herm}&
			\left\langle \begin{matrix} \phi X_v Z_w\herm \,,\, Z_v\herm X_w \end{matrix}\right\rangle
		\\[-0.75ex]\mapstoname{X_v \text{ meas.}}&
			\left\langle \begin{matrix} \phi X_v Z_w\herm \,,\, \omega^{-r} X_v \end{matrix}\right\rangle
		\;=\;
			\left\langle \omega^{-r} X_v \right\rangle \ox \left\langle \phi \omega^r Z_w\herm \right\rangle
		\\[-1ex]\mapstoname{X_w^{-r} \text{ corr.}}&
			\left\langle \omega^{-r} X_v \right\rangle \ox \left\langle \phi Z_w\herm \right\rangle	\;,
  \end{aligned}
\end{equation}
so that these operations transform $\phi X_v \mapsto \phi Z\herm_w$; and similarly,
\begin{equation}
  \begin{aligned}[b]
				\left\langle \phi Z_v \right\rangle
		\;\mapstoname{\text{prep.\ } \ket{+}_w}
			\left\langle \begin{matrix} \phi Z_v \,,\, X_w \end{matrix}\right\rangle
		\;\mapstoname{\cZ_{\!vw}\herm}&
			\left\langle \begin{matrix} \phi Z_v \,,\, Z_v\herm X_w \end{matrix}\right\rangle
		\;=\;
			\left\langle \begin{matrix} \phi Z_v \,,\, \phi X_w \end{matrix}\right\rangle
		\\[-1ex]\mapstoname{X_v \text{ meas.}}&
			\left\langle \begin{matrix} \omega^{-r} X_v \,,\, \phi X_w \end{matrix}\right\rangle
		\\[-1.5ex]\mapstoname{X_w^{-r} \text{ corr.}}&
			\left\langle \omega^{-r} X_v \right\rangle \ox \left\langle \phi X_w \right\rangle	\;,
  \end{aligned}
\end{equation}
so that we obtain $\phi Z_v \mapsto \phi X_w$.
\end{subequations}
Similarly, for any Weyl operator $W_{a,b}$ \cite[Definition~II]{Beaudrap2013b}, the operator $\phi W_{\!a,b}$ acting on $v$ will be transformed to a Weyl operator $\phi W_{\!-a,b}$ on $w$; the calculation is straightforward.
This implies (\emph{c.f.}~\cite[Eqn.~17]{Beaudrap2013b}) that aside from the teleportation from $v$ to $w$, the effect is an inverse Fourier transform of the state.

Thus, we may simulate the coding procedure of a node $V$ as described in Section~\ref{sec:coherentNetworkCoding} as follows.
Provided a collection of incoming qudits $a_1, \ldots, a_\ell$, we may prepare output qudits $b_1, \ldots, b_m$ by:
\begin{enumerate}
\item
	preparing output message qudits $b_1, \ldots, b_m$ and auxiliary qudits $b'_1, \ldots, b'_m$ in the state $\ket{+}$;

\item
	entangling the qudits $b_j$ and $b'_j$ by a $\cZ\herm$ operation, and performing $\cZ^{V_{jk}}$ operations between each pair of qudits $a_k$ and $b'_j$;

\item
	measuring each qudit $b'_j$ in the Fourier basis, obtaining some outcome $r_j$, and performing an $X^{-r_j}$ operation on the corresponding output qudit $b_j$.
\end{enumerate}
This describes a \MBQC\ procedure with inputs and outputs which we may illustrate by a \emph{geometry} (in the terminology of Ref.~\cite{Beau10,DKP06}) specifying the input and output qubits.
\begin{figure}
\begin{minipage}{0.59\textwidth}
  \caption{%
	Geometries of \MBQC\ procedures for a single node performing a transformation $V: \Z_d^\ell \to \Z_d^m$ of the standard basis.
	Incoming/outgoing message qudits are represented by blue circles; auxiliary qudits by black squares.
	\textbf{(a)}~The geometry associated to coding a single message qudit, simulating the right-hand circuit of Figure~\ref{fig:decompose-U_V}.
	Edges are labeled by their ``weights'', \ie~the necessary power of $\cZ$ in the procedure.
	As the qudits $a_k$ remain unmeasured, these are depicted as being outputs as well as inputs of this procedure.
	\textbf{(b)}~The geometry associated to the entire operation of a coding node, including measurement of the incoming message qudits.
	Edge weights between the qudits $a_k$ and $\alpha_j$ depend on the coding operation being simulated: if the coding operation being performed is sparse, many of these edge weights will be zero (corresponding to edges which should be omitted entirely).
	Only the qudits $b_j$ form the output of this procedure.
}
\label{fig:singlequditCodingMBQC}
\end{minipage}
\hfill
\begin{minipage}{0.45\textwidth}\centering
\begin{tikzpicture}[
										scale=0.7,
										to/.style={-stealth},
										line width=1pt, label distance=-3pt,
										site/.style={circle,fill=black!15!white,text=black!15!white, label distance=0pt},
										outtype/.style={circle,fill=blue!70!white, label distance=0pt},
										intype/.style={rectangle,fill=black, label distance=0pt}
										]
	\node (b'j) [intype, label={[label distance=-6pt]80:$b'_j$}] at (0.5,1) {};
	\node (bj)  [outtype, label={[label distance=-6pt]80:$b_j$}] at (2.5,1) {};

	\foreach \y/\n/\l/\p/\dx/\dy in {4/a1/1/0.6/0.4/0.1, 3/a2/2/0.5/-0.1/-0.45, 0/al/\ell/0.5/-0.1/0.35} {
		\node (\n) [outtype, label=90:$a_{\l}$] at (-1,\y) {};
		\draw (\n) -- node (\n') [pos=\p] {} (b'j);
		\node () at ($(\n') + (\dx,\dy)$) {\footnotesize$V_{j,\l}$};
		\draw [to] ($(\n) + (-1,0)$) -> (\n);
		\draw [to] (\n) -> ($(\n) + (1,0)$);
	}

	\node () at ($(a2)!0.55!(al)$) {\Large$\vdots$};

	\draw (b'j) -- node (b''j) {} (bj);
	\node () at ($(b''j) + (0,-0.35)$) {\footnotesize$-1$};	
	\draw [to] (bj) -> ($(bj) + (1,0)$);

	\node at (1,-0.5) {(a)};


	\foreach \ay/\an/\al in {-2/a1/1, -3/a2/2, -4/a3/3, -6/al/\ell} {
		\node (\an) [outtype, label={[label distance=-3pt]90:$a_{\al}$}] at (-1,\ay) {};
		\draw [to] ($(\an) + (-1,0)$) -- (\an);
	}

	\foreach \y/\bl in {-2.5/1, -3.5/2, -5.5/m} {
		\node (b'\bl) [intype, label={[label distance=-6pt]80:$b'_\bl$}] at (1,\y) {};
		\node (b\bl)  [outtype, label={[label distance=-6pt]80:$b_\bl$}] at (3,\y) {};
		\draw (b'\bl) -- node (b''\bl) {} (b\bl);
		\node () at ($(b''\bl) + (0,-0.25)$) {\footnotesize$-1$};	
		\draw [to] (b\bl) -- ($(b\bl) + (1,0)$);

		\foreach \ay/\an/\al in {4/a1/1, 3/a2/2, 2/a3/3, 0/al/\ell} {
			\draw (\an) -- (b'\bl);
		}
	}

	\node () at ($(a3)!0.35!(al)$) {\Large$\vdots$};
	\node () at ($(b'2)!0.45!(b'm)$) {\Large$\vdots$};
	\node () at ($(b2)!0.45!(bm)$) {\Large$\vdots$};

	\node at (1,-6.5) {(b)};

\end{tikzpicture}
\end{minipage}
\end{figure}
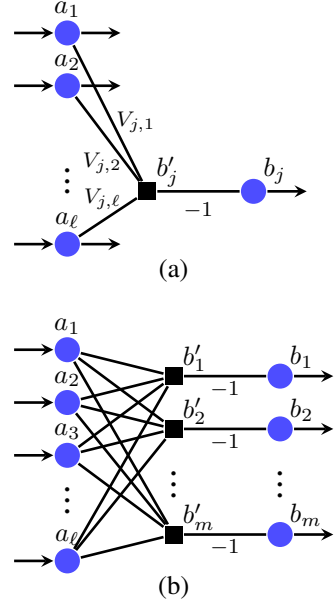
Figure~\ref{fig:singlequditCodingMBQC} presents geometries for the partial coding operation performed by $\tilde U_V$ as in Figure~\ref{fig:decompose-U_V}, and for the entire operation of a single coding node (including the eventual measurement of the input qubits): input qudits have arrows pointing inwards, and output qudits have arrows pointing outwards.

\subsection{\MBQC\ geometries to simulate entire network coding protocols}
\label{sec:geometriesWholeNetworks}

In the diagrammatic convention of this article, composition of \MBQC\ procedures may be represented by contracting the arrows between the outputs of earlier procedures and the inputs of later ones.
For \MBQC\ procedures to simulate the linear network codes, composing the geometries associated to each node yields a bipartite graph with a structure closely related to that of the coding network itself.
Specifically, one associates a qudit for the output qudits of the coding network, as well as for each incoming and outgoing message qudit at each node (with qudits at the outgoing links being the ``auxiliary'' qudits described above), and connecting them by a bipartite graph corresponding to the non-zero coefficients $V_{jk}$ of the coding node.
The edges of the coding network are replaced by \emph{undirected} edges with weights $-1$, corresponding to the entangling operations between the outgoing message qudits (which are either the inputs for some other node, or the outputs of the entire network).
The directionality of the communication links are represented by the order of the measurement and correction operations, as well as the classical communication involved in the correction subroutine.

\begin{figure}[t]
~\hfill
\begin{tikzpicture}[scale=0.75,-stealth, line width=1pt, label distance=-3pt, site/.style={circle,draw}]
	\node (null1) at (-0.5,4) {};
	\node (null2) at (-0.5,0) {};
	\node (S1) at (1,4) [site] {$S_1$};
	\node (S2) at (1,0) [site] {$S_2$};
	\node (V1) at (1.75,2) [site] {$V_1$};
	\node (V2) at (3.75,2) [site] {$V_2$};
	\node (T1) at (4.5,4) [site] {$T_1$};
	\node (T2) at (4.5,0) [site] {$T_2$};
	\node (null3) at (6,4) {};
	\node (null4) at (6,0) {};

	\foreach \s/\t/\r/\l/\a in {null1/S1/0.35/s_1/90, S1/T1/0.475/m_3/90, T1/null3/0.55/t_1/90, null2/S2/0.35/s_2/-90, S2/T2/0.475/m_5/90, T2/null4/0.55/t_2/-90, S1/V1/0.45/m_1/180, S2/V1/0.45/m_2/180, V1/V2/0.45/m_4/90, V2/T1/0.45/m_6/0, V2/T2/0.45/m_7/0} {%
		\draw	(\s) -> (\t);
		\node (\l) [fill=blue!70!white, circle, label=\a:$\l$] at ($(\s)!\r!(\t)$) {};
	}

	\node at (2.7,-1) {\textup{(a)}};
\end{tikzpicture}
\hfill
\begin{tikzpicture}[scale=0.75, to/.style={-stealth}, line width=1pt, label distance=-3pt, site/.style={circle,fill=black!15!white,text=black!15!white}, intype/.style={circle,fill=blue!70!white}, outtype/.style={rectangle,fill=black}]
	\node (null1) at (-0.5,4) {};
	\node (null2) at (-0.5,0) {};
	\node (S1) at (1,4) [site] {$S_1$};
	\node (S2) at (1,0) [site] {$S_2$};
	\node (V1) at (1.75,2) [site] {$V_1$};
	\node (V2) at (3.75,2) [site] {$V_2$};
	\node (T1) at (4.5,4) [site] {$T_1$};
	\node (T2) at (4.5,0) [site] {$T_2$};
	\node (null3) at (6.5,4) {};
	\node (null4) at (6.5,0) {};

	\foreach \s/\t/\l/\a/\d in {null1/S1/s_1/90/0, S1/T1/m_3/115/1, T1/null3/t_1/90/1, null2/S2/s_2/-90/0, S2/T2/m_5/-115/1, T2/null4/t_2/-90/1, S1/V1/m_1/180/1, S2/V1/m_2/180/1, V1/V2/m_4/90/18, V2/T1/m_6/-180/1, V2/T2/m_7/180/1} {%
		\ifnum\d<2
			\node (\l) 	[intype,label=\a:$\l$] at ($(\t)!14pt!(\s)$) {};
		\else
			\node (\l) 	[intype,label=\a:$\l\mspace{\d mu}$] at ($(\t)!14pt!(\s)$) {};
		\fi
		\ifnum\d>0
			\node (\l') [outtype] at ($(\s)!14pt!(\t)$) {};
			\draw (\l) -- (\l');
		\fi
	}

	\foreach \s/\t/\l/\a/\d in {T1/null3/t_1/90/8, T2/null4/t_2/-90/8, V1/V2/m_4/-90/18} {%
		\ifnum\d<2
			\node (\l') 	[outtype,label=\a:$\l'$] at ($(\s)!14pt!(\t)$) {};
		\else
			\node (\l') 	[outtype,label=\a:$\mspace{\d mu}\l'$] at ($(\s)!14pt!(\t)$) {};
		\fi
	}
	
	\draw [to] 	(null1) -> (s_1);
	\draw [to]	(null2) -> (s_2);
	\draw [to]	(t_1) -> ($(null3)+(0.25,0)$);
	\draw [to]	(t_2) -> ($(null4)+(0.25,0)$);

	\foreach \v in {m_1, m_2, m_3, m_4, m_5, m_6, m_7, t_1, t_2}
		\draw (\v') -- (\v);

	\foreach \s/\i/\t in {s_1/S1/m_1, s_1/S1/m_3, m_1/V1/m_4, m_2/V1/m_4, s_2/S2/m_2, s_2/S2/m_5, m_3/T1/t_1, m_4/V2/m_6, m_4/V2/m_7, m_5/T2/t_2, m_6/T1/t_1, m_7/T2/t_2}
		\draw (\s) .. controls ($(\i)!5pt!(\s)$) and ($(\i)!5pt!(\t)$) .. (\t');

	\node at (2.75,-1) {\textup{(b)}};	
\end{tikzpicture}
\hfill~\\[3ex]

~\hfill
\begin{tikzpicture}[yscale=0.75, xscale=0.5, to/.style={-stealth}, line width=1pt, label distance=-3pt]
	\node (null1) at (-1,4) {};		\node (null2) at (-1,0) {};
	\node (null3) at (5,4) {};		\node (null4) at (5,0) {};

	\foreach \l/\x/\y/\a in {s_1/0/4/90, s_2/0/0/-90, m_1/0/2.66/180, m_2/0/1.33/180, m_3/2/4/90, m_4/3/2/0, m_5/2/0/-90, m_6/3/3.33/0, m_7/3/0.66/0, t_1/4/4/90, t_2/4/0/-90} {
		\node (\l) [circle, fill=blue!70!white, label=\a:$\l$] at (\x,\y) {};
	}

\foreach \l/\x/\y/\a in {m_4'/0/2/180, t_1'/3/4/90, t_2'/3/0/-90} {
		\node (\l) [rectangle, fill=black, label=\a:$\l$] at (\x,\y) {};
	}

	\foreach \l/\x/\y in {m_1'/0/3.33, m_2'/0/0.66, m_3'/1/4, m_5'/1/0, m_6'/3/2.66, m_7'/3/1.33} {
		\node (\l) [rectangle, fill=black] at (\x,\y) {};
	}
 
	\draw [to] ($(null1)+(-0.25,0)$) -- (s_1);
	\draw [to] ($(null2)+(-0.25,0)$) -- (s_2);
	\draw [to] (t_1) -- ($(null3)+(0.25,0)$);
	\draw [to] (t_2) -- ($(null4)+(0.25,0)$);

	\draw (s_1) -- (m_1') -- (m_1) -- (m_4');
	\draw (s_2) -- (m_2') -- (m_2) -- (m_4');
	\draw (s_1) -- (m_3') -- (m_3) -- (t_1') -- (t_1);
	\draw (s_2) -- (m_5') -- (m_5) -- (t_2') -- (t_2);
	\draw (m_4') -- (m_4);
	\draw (m_4) -- (m_6') -- (m_6) -- (t_1');
	\draw (m_4) -- (m_7') -- (m_7) -- (t_2');

	\node at (2,-1.25) {\textup{(c)}};	
\end{tikzpicture}
\hfill
\begin{tikzpicture}[scale=0.5, to/.style={-stealth}, line width=1pt, label distance=-3pt]
	\node (null1) at (-1,4) {};		\node (null2) at (-1,0) {};
	\node (null3) at (5,4) {};		\node (null4) at (5,0) {};

	\foreach \l/\x/\y/\a in {s_1/0/4/90, s_2/0/0/-90, m_4/2/2/0, t_1/4/4/90, t_2/4/0/-90} {
		\node (\l) [circle, fill=blue!70!white, label=\a:$\l$] at (\x,\y) {};
	}

	\foreach \l/\x/\y/\a in {m_4'/0/2/180, t_1'/2/4/90, t_2'/2/0/-90} {
		\node (\l) [rectangle, fill=black, label=\a:$\l$] at (\x,\y) {};
	}
 
	\draw [to] ($(null1)+(-0.25,0)$) -- (s_1);
	\draw [to] ($(null2)+(-0.25,0)$) -- (s_2);
	\draw [to] (t_1) -- ($(null3)+(0.25,0)$);
	\draw [to] (t_2) -- ($(null4)+(0.25,0)$);

	\draw (s_1) -- (m_4');
	\draw (s_2) -- (m_4');
	\draw (s_1) -- (t_1') -- (t_1);
	\draw (s_2) -- (t_2') -- (t_2);
	\draw (m_4') -- (m_4);
	\draw (m_4) -- (t_1');
	\draw (m_4) -- (t_2');

	\node at (2,-2.875) {\textup{(d)}};	
\end{tikzpicture}
\hfill~
\vspace{-1ex}
\caption{%
	Construction of a \MBQC\ geometry for a procedure simulating a coding protocol for the $2$-pair problem 
	on \textbf{(a)}~the butterfly network, shown with message qudits for each communication link.
	\textbf{(b)}~The~graph obtained by substituting each coding node, with the geometry for the corresponding \MBQC\ procedure.
	This is derived by adding vertices for ``auxiliary'' qudits (black squares) for each output message qudit, and associating each ``auxiliary--output'' pair to an outbound network link.
	Edges represent powers of 
	$\cZ$ operations, which are used for single-qudit teleportation along the network links. 
	The input and output message qudits of the linear code become the source and target subsystems of the \MBQC\ procedure.
	\textbf{(c)}~The~same geometry, presented in grid formation.
	\textbf{(d)}~The geometry of a \MBQC\ procedure (\emph{c.f.}~Ref.~\cite[Figure~7]{BKMP07}) for the \textsc{swap} operation.
%
%
%
}
\label{fig:butterflyToMBQC}
\end{figure}
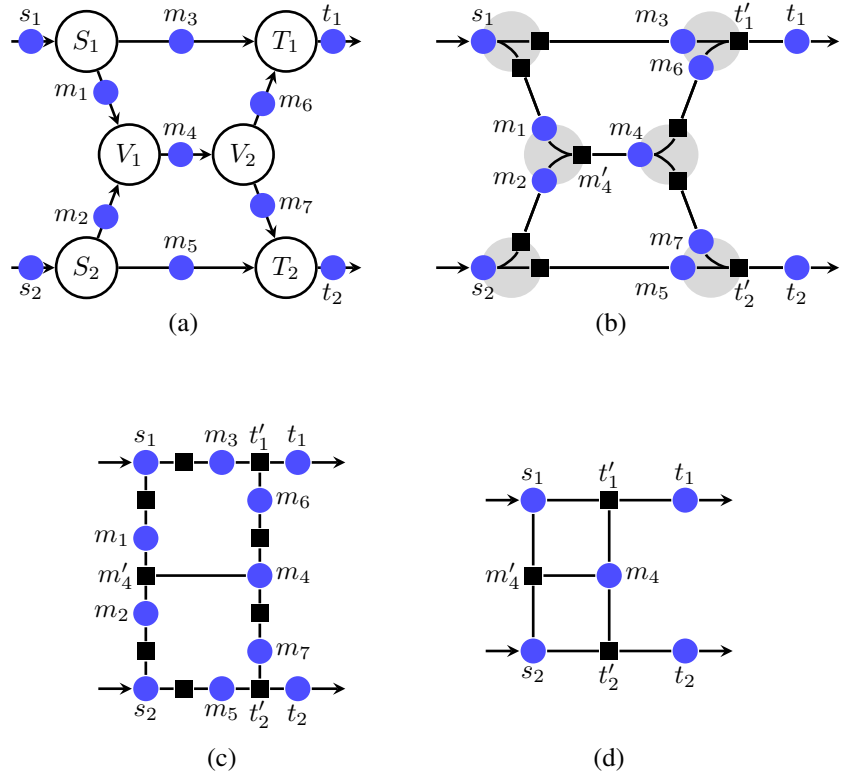

As an example, we illustrate this construction in Figure~\ref{fig:butterflyToMBQC} for procedure for the two-pair problem performing a \textsc{swap} operation on two qudits (\eg~in which we use the coding operations $S_1 = S_2 = V_2 = \mbox{$[\;1\;\;1\;]$}\trans$ and $V_1 = T_1 = T_2 = \mbox{$[\,{-}1\;{-}1\,]$}$).
 
As every measurement involved is performed in the Fourier basis (equivalently: the eigenbasis of the $X$ operator), the only information which this graphical representation omits are the order in which the measurements occur, and the correction procedures, which we consider next.

\subsection{Measurement and communication of outcomes}

The corrections required to use $X$ measurements to simulate projection onto $\ket{+}$ may be performed in two natural ways, corresponding to the protocols of Refs.~\cite{KLGNR09b} and~\cite{KLGNR10} respectively.

\subsubsection{Free classical communication}

In a setting as in Ref.~\cite{KLGNR09b} where classical communication is free, all corrections may be deferred to the target nodes of the coding network, which prepare the output qudits.
This is a natural approach for simulating the network code as a \MBQC\ procedure: in measurement-based computation, it is conventional to simulate CP maps in such a way that the output qudits are the only qudits on which unitary correction operations are performed.
As in Ref.~\cite{KLGNR09b}, successful projection onto the $\ket{+}$ state (or a ``0'' outcome of a $X$ measurement) is the ideal case; it then suffices to determine how the errors (or \emph{byproduct operations} in the terminology of Ref.~\cite{RBB03}) propagate to the output qudits, in order to correct them.
We describe this in terms of communication directly to the targets, as well as some amount of communication within the coding network.

When simulating the coding procedure at each node using auxiliary qudits, measuring those auxiliary qudits introduces an additional source of error: if the correction is not immediately performed on the outgoing message qudits, this induces additional phase errors.
Commuting an $X_{b_j}^{-r}$ operation past an entangling operation $\cZ^{U_{ij}}_{b_j c'_i}$, where $c'_i$ is an auxiliary qudit for a subsequent node performing a coding operation $U$, yields an error operation $X_{b_j}^{-r} Z_{c'_i}^{-\smash{r U_{ij}}}$.
The operation $\smash{X_{b_j}^{-r}}$ does not affect the outcome of the measurement on $b_j$, as the states $\ket{\omega_r}$ are eigenvectors of $X$.
The $Z$ error on $c'_i$ induced by postponing the correction on $b_j$ is significant, but we may account for this error by classical post-processing of the measurement result $r'$ on $c'_i$ itself.
Let $\tilde r = r U_{ij}$ for the sake of brevity: because $X Z^{- \tilde r} \propto = \omega^r Z^{-\tilde r} X$, we may account for an uncorrected $Z^{-\tilde r}$ operation on $c'_i$ by performing an $X$ measurement, obtaining some outcome $r_0'$, and then subtracting $\tilde r$ from that outcome to obtain an adjusted outcome $r' = r'_0 - \tilde r$ for future corrections.

More generally, $c'_i$ will accumulate uncorrected $Z$ errors arising from the uncorrected $X$ errors on each of the input messages on which it depends.
If those input qubits $b_j$ have errors $X^{-r_j}$ associated with them, these collectively induce an error
\begin{equation}
  Z^{-(r_1 U_{i1} + r_2 U_{i2} + \cdots)} = Z^{-\unit_i \cdot U\vec r}
\end{equation}
on $c'_i$.
We may simulate this correction after the $Z$ measurement by subtracting $\tilde r = \unit_i \cdot U\vec r$ from the measurement outcome $r'_0$, yielding $r' = r'_0 \,-\, \unit_i \cdot U\vec r$.
By propagating the results of the auxiliary qudit measurements forward through the coding network, subsequent coding nodes may locally adapt the measurement outcomes in order to simulate the correction of errors on their own auxiliary qudits, allowing the target nodes to perform the necessary $X$ corrections on the output qudits of the network.
Alternatively, all of the results may be transmitted directly to the target nodes, which can simulate this sequential adaptation of measurement outcomes themselves.

For a coding network performing an injective transformation $M: \Z_d^{\mathscr S} \to \Z_d^{\mathscr T}$, the phase errors induced by measurement of the message qudits may be corrected in the manner described in Ref.~\cite{KLGNR09b}.
Without loss of generality, we may suppose that the agents at each network coding node prepare their auxiliary and message qudits, and all nodes except the target nodes communicate their outgoing messages to their recipients.
Afterwards, they measure their auxiliary nodes in some order consistent with the topological ordering of the network, and similarly communicate the outcomes forward, allowing subsequent nodes to adjust their auxiliary measurement outcomes, and allowing target nodes to perform what $X$ corrections are necessary on the output qudits.
The remaining measurement operations and classical messages are identical to those of Ref.~\cite{KLGNR09b}, in which it does not matter if nodes transmit outgoing message qudits before they measure incoming message qudits.

For the sake of completeness, we sketch an inductive approach to the $Z$ correction protocol of the target nodes in this setting.
Let $A$ be a left-inverse of $M$, and consider an input state $\ket{\psi}$ to the coding network, expressed as
\begin{equation}
		\ket{\psi}
	\;=\;	
		\sum_{\vec x \in \Z_d^{\mathscr S}} u_{\vec x} \ket{\vec x}
	\;\,=\!\!\!\!
		\sum_{\vec y \in \img(M)} \!\!\!\!u_{A \vec y} \ket{A \vec y}.
\end{equation}
The state obtained after performing the preparation and entanglement phases of the \MBQC\ procedure, and after performing the auxiliary qudit measurements and $X$ corrections on the output qudits, is exactly a state of the form in Eqn.~\eqref{eqn:entangledQuantumNetworkCode}, of the form
\begin{equation}
	\begin{aligned}[b]
		\ket{\Psi}
	\,=\!\!\!\!\!\!
		\sum_{\vec y \in \img(M)} \!\!\!\!\!\!\!u_{A \vec y} \ket{A \vec y}_{\!\mathsf S} &\ox \ket{M A \vec y}_{\!\mathsf T} \ox \ket{H A \vec y}_{\!\text{rest}}\,
	\\[-2ex]&=\!\!\!\!\!\!
		\sum_{\vec y \in \img(M)} \!\!\!\!\!\!\!u_{A \vec y} \ket{A \vec y}_{\!\mathsf S} \ox \ket{\vec y}_{\!\mathsf T} \ox \ket{H A \vec y}_{\!\text{rest}}
	\end{aligned}
\end{equation}
for some linear map $H$.
(The latter equality holds because for any $\vec y = M \vec x$, we have $MA\vec y = MAM\vec x = \vec y$.)
Indeed, the distinction between the input qudits $\mathsf S$ and the other non-target qudits is unimportant: we may subsume the linear map $A$ on the standard basis of $\mathscr S$ and the map $HA$ on the standard basis of the other qudits into a map
\begin{equation}
  K \;=\;	\left[\;\:\begin{matrix} \!\!\! A \!\!\! \\\hline\\[-2ex] \!\!\! HA \!\!\! \end{matrix}\;\:\right]
\end{equation}
where the upper rows correspond to indices in $\mathsf S$, and the lower rows to the other non-output qudits.
We may then write
\begin{equation}
	\begin{aligned}[b]
		\ket{\Psi}\;
	\,&=\!\!\!\!
		\sum_{\vec y \in \img(M)} \!\!\!\!u_{A \vec y} \ket{\vec y}_{\!\mathsf T} \ox \ket{K \vec y}_{\!\Omega \setminus \mathsf T}\;.
	\end{aligned}
\end{equation}
We may isolate any non-output qudit $u \in \Omega \setminus T$.
Let $\Omega' = \Omega \setminus \{u\}$, and consider another decomposition
\begin{equation}
  K \;=\;	\left[\;\:\begin{matrix} \!\!\! \bm \kappa_u\trans  \!\!\! \\\hline\\[-2ex] \!\!\! K' \!\!\! \end{matrix}\;\:\right]
\end{equation}
where the upper row corresponds to the index for the qudit $u$ and contains a row-vector $\bm \kappa_u\trans$\,, and $K'$ corresponds to all of the other non-output qudits; we may then once more re-write
\begin{equation}
	\begin{aligned}[b]
		\ket{\Psi}\;
	&=\!\!\!\!\!
 		\sum_{\vec y \in \img(M)} \!\!\!\!\!u_{A \vec y} \ket{\vec y}_{\!\mathsf T} \; \ket{\bm \kappa_u \!\cdot \vec y}_{\!u} \ox \ket{K' \vec y}_{\!\Omega' \setminus \mathsf T }.
	\end{aligned}
\end{equation}
Measuring $u$ in the Fourier basis and obtaining the outcome $r$, the resulting state on the remaining qudits is
\begin{equation}
		\ket{\Psi'}\;
	=\!\!\!\!\!
		\sum_{\vec y \in \img(M)} \!\!\!\!\!u_{A \vec y} \, \omega^{-r(\bm \kappa_u \!\!\;\cdot\!\: \vec y)} \ket{\vec y}_{\!\mathsf T} \; \ket{K' \vec y}_{\!\Omega' \setminus \mathsf T},
\end{equation}
following Eqn.~\eqref{eqn:measOnEntangledQuantumNetworkCode}.
If the outcome $r$ is transmitted to the target nodes, and who know the value of $\bm \kappa_u$, they may simply compute $\bm \sigma  := r \bm \kappa_u$ and collectively perform $Z^{\bm \sigma} = Z_{t_1}^{\sigma_1} Z_{t_2}^{\sigma_2} \cdots$ on the qudits of $\mathsf T$, thereby obtaining
\begin{equation}
		\ket{\Psi''}\;
	=\!\!\!\!\!
		\sum_{\vec y \in \img(M)} \!\!\!\!\!u_{A \vec y} \ket{\vec y}_{\!\mathsf T} \; \ket{K' \vec y}_{\!\Omega' \setminus \mathsf T},
\end{equation}
which is again a state of the same form as in Eqn.~\eqref{eqn:entangledQuantumNetworkCode}, on one fewer qudits.
By induction, we may measure each of the qudits of $\Omega \setminus \mathsf T$ in any order (or simultaneously), and transmit them to the target nodes, which then make the appropriate $Z$ corrections to obtain the state
\begin{equation}
		\bigl\lvert\Psi\sur{n}\bigr\rangle
	\;=\!\!\!
		\sum_{\vec y \in \img(M)} \!\!\!\!\!u_{A \vec y} \ket{\vec y}_{\!\mathsf T}
	\;=
		\sum_{\vec x \in \Z_d^{\mathscr S}} \!u_{\vec x} \ket{M \vec x}_{\!\mathsf T}	\;.
\end{equation}
In summary, provided free classical communication to the targets and within the coding network, all measurements may be performed simultaneously, with the results of the measurement of incoming messages being transmitted directly to the targets to perform $Z$ corrections on the output qudits.
Measurement results of the auxiliary qudits may be communicated along the coding network, and used to adapt the outcomes of subsequent measurements, culminating in measurement information useful to the target nodes to perform $X$ corrections on the output qudits.

\subsubsection{Constrained classical communication}

In the setting of Ref.~\cite{KLGNR10}, we attempt to reduce the amount of classical communication which takes place outside of the network (but allowing messages to pass in either direction).
To this end, we allow the source nodes and the intermediate nodes of the network to perform $Z$ corrections.
The way in which these corrections are performed follows from \textbf{(a)}~the description of how $X$ corrections may be simulated in the setting of ``free'' classical communication, as this already can be performed only with communication within the coding network; and \textbf{(b)}~the phase correction procedure of Ref.~\cite{KLGNR10} which was outlined in Section~\ref{sec:coherentNetworkCoding}.
These corrections may be performed as follows:
\begin{itemize}
\item 
	All auxiliary qudits may be measured simultaneously, and their outcomes propagated forward through the network, as in the previous section.
	Alternatively, one may instead perform $X$ correction operations for the auxiliary qudits at each node: this imposes an order on the measurement of the auxiliary qudits which is consistent with the topological order of the network, so that each node may use the measurement outcomes for preceding auxiliary qudits when correcting its own auxiliary qudits.
\item
	The measurement of each node's incoming message qudits must be performed in an order opposite to the topological order of the coding network, in order to allow the node which sent each message qudit to perform the necessary corrections involving its own incoming message qudits.
\end{itemize}
From this, one may derive schedules for measuring each qudit in the network, and for communicating classical messages forward or backward through the network to allow the necessary $X$ or $Z$ corrections.

For the correction of phases induced by measurement of the input qubits of the source, following As in Section~\ref{sec:coherentNetworkCoding}, whether the corrections arising from the measurement of the input qudits managed by the source nodes can be corrected without communicating outside of the network, may depend on the transformation which the network performs.
For any linear transformation $M$ for which $M\trans B M = \idop$ for some block-diagonal $B$ acting on blocks of qudits held by target nodes --- \eg~for permutation matrices $M$ --- classical network coding of of the outcomes of measuring the inputs of the source nodes will suffice.

\subsection{Overview of the \MBQC\ construction}

The above construction rests on the fact that the protocol of Ref.~\cite{KLGNR09b} is unaffected if the measurements are deferred until each node sends its messages.
(The protocol of Ref.~\cite{KLGNR10} in fact requires this modification.)
The result of doing so causes these protocols to give rise to large distributed entangled states, on which local measurements are performed to simulate projection onto the $\ket{+}$ state.
In this sense, these protocols are literally quantum computation by measurements; the modifications described in this Section --- namely, replacement of $\cX$ operations by $\cZ$ operations, introduction and measurement of auxiliary qudits in order to make the previous modification possible, and communication of the results of measuring auxiliary qudits --- are straightforward modifications which demonstrate that they are effectively computations in the one-way \MBQC\ model of Refs.~\cite{RBB03,DKP06}.

The \MBQC\ procedures which result from these transformations have comparable complexity to the original protocols of Refs.~\cite{KLGNR09b,KLGNR10}, differing essentially only in the various operations performed on the auxiliary qudits, as well as the communication and transformation of their measurement outcomes.
For a coding network with $k$ input messages, $\ell$ output messages, and $m$ internal communication links, the total number of qudits involved in the \MBQC\ procedure is easily verified to be $k + 2\ell + 2m$, following Section~\ref{sec:geometriesWholeNetworks}.
The number of entangling operations involved for each node (disregarding exponents) is simply the same as the number of $\cX$ operations involved in simulating $\tilde U_V$, plus twice the out-degree (involved in entangling the auxiliary and outgoing message qudits for the node).
Thus there are exactly $2(m+\ell)$ more entangling operations, in the form of $\cZ$ operations, in the \MBQC\ protocol than there are $\cX$ operations in the original presentation of the protocols in Refs.~\cite{KLGNR09b,KLGNR10}.
There are also exactly $2(m+\ell)$ additional classical messages sent in the \MBQC\ protocol, either directly to the targets or entirely within the network, again as a result of measuring the auxiliary qudits.

\section{Open questions}

In this article, we have illustrated the way in which classically-assisted quantum linear network coding over $\Z_d$ as described by Kobayashi \etal~\cite{KLGNR09b,KLGNR10} is in effect an instance of measurement-based computation in the one-way model~\cite{RBB03,DKP06}, in particular using measurements only in the Fourier basis (the eigenbasis of the $X$ cyclic shift operator on $d$-dimensional qudits).
While not explicitly presented as an instance of \MBQC, the differences between the protocols of Refs.~\cite{KLGNR09b,KLGNR10} and one-way measurement-based procedures are straightforward, and involve no substantial differences in \eg~the amount of classical communication required.
We may ask to what extent these results (particularly the bounds on classical communication outside of the network) hold for classically assisted \emph{non-linear} quantum codes as well.

While the \MBQC\ model is sometimes described as a distributed model of computation, little emphasis has been placed on the communication cost of \MBQC\ computation.
A common presentation (\eg~as in Refs.~\cite{BFK09,AB09}) is that measurement results are recorded by an effectively delocalized classical control, which receives messages containing measurement outcomes from one or more agents which manage individual qudits, and which responds with instructions of how to perform subsequent measurements.
Bounding the communication requirements of a \MBQC\ procedure, to eliminate the need of a delocalised control center, may be necessary to realize the reduction in the computational depth of a \MBQC\ procedure (one of the theoretical selling points of the \MBQC\ model~\cite{RBB03}).

As network coding subsumes constant-depth distributed computation, we may interpret these results as recommending measurement-based computation as a framework for analyzing multiparty communication protocols, as we have suggested in the introduction.
We may also consider this as an alternative means of approaching the problem of assigning semantics to measurement-based computations, a problem of some interest in models of quantum computation~\cite{DKP06,Beau10,Duncan-2012,SGK-2011}.
Specifically: rather than interpreting a measurement-based procedure as a quantum circuit with some potentially exotic features (such as closed time-like curves~\cite{SGK-2011}), we may interpret pieces of measurement-based computations as coherently simulating transformations of the standard basis on several qudits at once.
Such simple semantics is likely to prove useful to any programme to find novel ways of using measurement-based computation as a medium in which to develop algorithms (see~Ref.~\cite{BDKR08}).

As a final open question, we ask whether a converse to our results hold, the form of a classical simulation algorithm for certain measurement-based computations by linear network codes.
This article shows that (a coherent quantum simulation of) a classical linear network code is in effect a measurement-based procedure which performs only $X$-eigenbasis measurements, on a graph state with similar structure to the coding network.
This is a special case of an efficiently simulatable class of computations: the unitary transformations realized by \MBQC\ procedures performing only Pauli-eigenbasis measurements are \emph{Clifford group operations},\footnote{%
	This is well-known for qubits~\cite{BB06}; on qudits it follows from how stabilizer states are transformed under measurements, see Ref.~\cite{Beaudrap2013b}.
}
which can be simulated \eg~on standard basis states by linear transformations on a cyclic ring~\cite{Beaudrap2013b}.
This raises the question: is there a sense in which a \MBQC\ procedure on a graph $G$, which implements unitary a transformation using only measurements in a Pauli eigenbasis (or only the $X$-eigenbasis) and Pauli corrections, can be ``locally'' simulated by a classical linear code --- in such a way that the expectation value of any observable on a single given qudit can be evaluated from information available at a corresponding target node --- on a network similar to $G$?

\subsubsection*{Acknowledgements.}

This work was done in part while MR was with NEC Laboratories America, and NdB was at the University of Cambridge with support from the EC project QCS.
NdB would like to thank Peter H{\o}yer for helpful comments at the beginning of this research.

\bibliography{network-coding-lipics}
\end{document}